\documentclass[aps,prl,amsmath,amssymb,reprint,superscriptaddress,twocolumn,showpacs]{revtex4-1}

 \usepackage{graphicx}
 \usepackage{amsmath}
 \usepackage{amsfonts}
 \usepackage{amssymb}
 \usepackage{pstricks}
 \usepackage{psfrag}
 \usepackage{epsfig}
 \usepackage{changes}
 \usepackage[ansinew]{inputenc}
\usepackage{mathptmx}
\usepackage{helvet}
\usepackage{textcomp}
\usepackage{ulem}

\begin{document}


\title{Two-phase crystallisation in a carpet of inertial spinners}



\author{Zaiyi Shen}
\affiliation{Univ. Bordeaux, CNRS, LOMA, UMR 5798, F-33400 Talence, France}
\author{Juho S. Lintuvuori}
\email[]{juho.lintuvuori@u-bordeaux.fr}
\affiliation{Univ. Bordeaux, CNRS, LOMA, UMR 5798, F-33400 Talence, France}


\date{\today}

\begin{abstract}
We study the dynamics of torque driven spherical spinners settled on a surface, and demonstrate that hydrodynamic interactions at finite Reynolds numbers can lead to a concentration dependent and non-uniform crystallisation.
At semi-dilute concentrations, we observe a rapid formation of a uniform hexagonal structure in the spinner monolayer. We attribute this to repulsive hydrodynamic interactions created by the secondary flow of the spinning particles. Increasing the surface coverage leads to a state with two co-existing spinner densities. The uniform hexagonal structure deviates into a high density crystalline structure surrounded by a continuous lower density hexatically ordered state.
We show that this phase separation occurs due to a non-monotonic hydrodynamic repulsion, arising from a concentration dependent spinning frequency.
\end{abstract}


\maketitle


\paragraph{Introduction:}
Recently, active systems considering rotational degrees of freedom have emerged as an important part of out-of-equilibrium materials~\cite{riedel2005self,driscoll2019leveraging,driscoll2017unstable,karani2019tuning,kokot2018manipulation,kaiser2017flocking,martinez2018emergent,bricard2013emergence,drescher2009dancing,petroff2015fast,van2016spatiotemporal,banerjee2017odd,soni2019odd,jaeger2013dynamics}.  In the case of field-actuated colloidal particles~\cite{bricard2013emergence, driscoll2017unstable,karani2019tuning,kokot2018manipulation,kaiser2017flocking,martinez2018emergent}, the rotational and translational motion are coupled. The flow created by the rotating objects in the presence of a boundary can lead to a translational motion. The rolling objects have been observed to form rotating clusters~\cite{karani2019tuning} and the hydrodynamic coupling between the rollers have been attributed to a formation of flocking states~\cite{bricard2013emergence} as well as to a fingering instability~\cite{driscoll2017unstable}. While purely rotational dynamics has been linked to the formation of hexagonal crystal by fast spinning bacteria~\cite{petroff2015fast} as well as to the emergence of edge currents ~\cite{van2016spatiotemporal} and odd viscosity~\cite{banerjee2017odd} in dry spinner materials.

Another example, where the individual dynamics is purely rotational, is provided by torque driven particles suspended in a fluid.
Previous work has predicted a phase separation of binary mixture of counter-rotating spheres in a monolayer at the Stokes' limit~\cite{yeo2015collective}.
In the absence of inertia, an individual spinning sphere creates a rotational flow field with only an azimuthal component~\cite{climent2007dynamic,fily2012cooperative}. At higher volume fractions this enables the particles to explore different states and leads to chaotic particle trajectories~\cite{lushi2015periodic}.
At hexagonally symmetrical arrangement of the particles the mutual (azimuthal) flow fields cancel, rendering the structure marginally stable~\cite{lenz2003membranes,lenz2004membranes} and recently it has been shown that combining the mixing arising from the rotational flow with a steric repulsion can lead to a fast crystallisation  at zero Reynolds number ($\mathrm{Re}$) limit~\cite{oppenheimer2019rotating}.

When the rotational $\mathrm{Re}$ is increased, {\color{black}  inertial effects become important.}
Co-rotating disks on a gas-liquid interface have been observed to form hexagonal arrangements due to an interplay between  repulsive far-field Magnus forces and a magnetic attraction~\cite{grzybowski2000dynamic,grzybowski2001dynamic}, while simulations have predicted both attractive and repulsive hydrodynamic interactions between co-spinning disks at finite Reynolds numbers in strictly 2-dimensions (2D)~\cite{goto2015purely}.

In 3-dimensions (3D) and at $\mathrm{Re}\sim 1$ a single spinning spherical particle creates an additional flow, which includes both radial and polar components~\cite{climent2007dynamic,collins1955steady,bickley1938lxv} which are missing in the 2D case. This secondary flow has been attributed to the repulsion between a spinner pair~\cite{climent2007dynamic,aragones2016elasticity,steimel2016emergent} and to the attraction {\color{black} of a single spinner} towards a no-slip wall along the spinning axis~\cite{liu2010wall}. {\color{black} At higher volume fractions, the secondary flow is expected to lead more intricate particle dynamics and, for example, the stabilisation of spinner vortices in 3D space has been predicted~\cite{shen2020hydrodynamic}.}

\begin{figure}
\centering
\includegraphics[width=1\columnwidth]{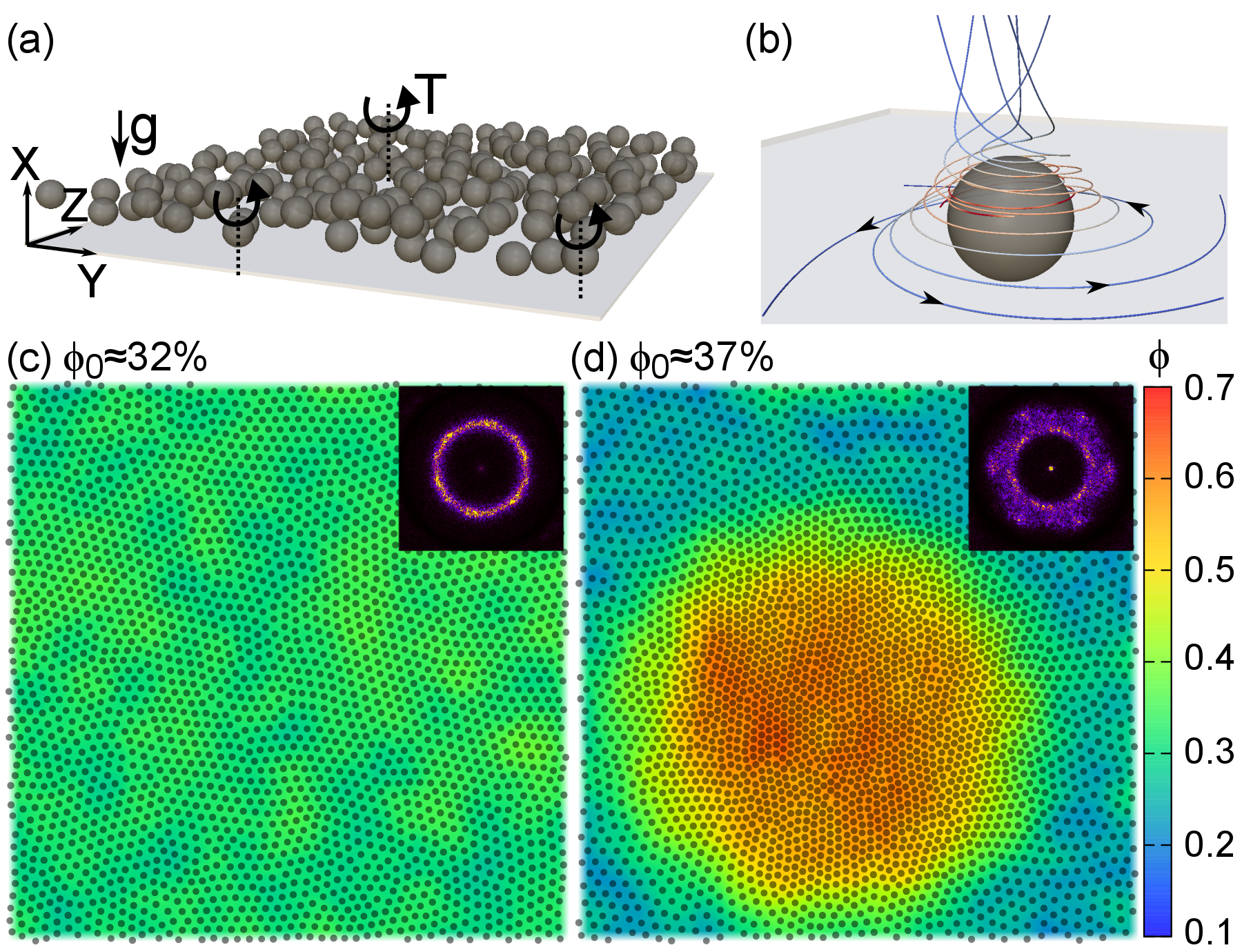}
\caption{(a) A schematic showing the simulation system. A constant torque $T$ is subjected to each particle sedimented on a flat wall. (b) The streamlines showing the flow field created by a rotating spherical particle with $\mathrm{Re}\approx 10$. (c) A hexagonal structure of the particles for an area fraction $\phi_0 \approx 32\%$. (d) Phase separation to particle dense and dilute regions for $\phi_0 \approx 37\%$ . The colour maps show the local area fractions. The insets in (c) and (d) show the structure factors with the characteristics of a hexatic order in both cases.
\label{config}}
\end{figure}

{\color{black}Here we study the inertial hydrodynamics of spinners at high concentrations. The system consists of spherical spinning particles near a no-slip surface and includes both the effects of inertia and the 3D flow fields.}
The particles are subjected to a constant torque in the wall normal and to a weak gravity towards the surface (Fig.~\ref{config}(a)). At the steady state the spinners form a monolayer.
Starting from random initial positions above the surface, we observe a rapid formation of  hexatic order  at semi-dilute area coverages (Fig.~\ref{config}(c)). In the absence of thermal effects, the crystallisation arises from an interplay between the hydrodynamic mixing from the azimuthal flow fields and the repulsion from the secondary flow. When the overall area fraction is increased, we find a spontaneous condensation of high spinner density area surrounded by a lower density hexagonal structure of the spinners (Fig.~\ref{config}(d)). We show that this phase separation is due to particle concentration dependent hydrodynamic repulsion {\color{black} - the spinning frequency decreases with increasing concentration, leading to a reduction of the hydrodynamic repulsion at high particle densities.}

\paragraph{Methods and simulation set-up:}
We employ a lattice Boltzmann method (LBM) to solve the dynamics of the system. The fluid-particle interaction is achieved by bounce-back on links method~\cite{ladd1,ladd2,nguyen2002lubrication}, which gives rise to a no-slip boundary condition on the particle surface.
A very short range repulsion between the solid boundaries is applied in order to avoid particle-particle and particle-wall overlaps~\cite{shen2018hydrodynamic,shen2019hydrodynamic}. We consider a system where many spherical particles are subjected to a torque and a weak gravity above a flat surface (Fig.~\ref{config}(a)).
We study the system at finite rotational Reynolds numbers ($\mathrm{Re}>1$), which measure the ratio between inertial and viscous forces $\mathrm{Re}=\rho \omega R^2/\mu$, where $\omega$ and $R$ are the rotational frequency and radius of a particle, and $\rho$ and $\mu$ are the density and viscosity of the fluid~\cite{SupMat}. The particles are driven by a constant torque $T$, which leads to the spinning motion of the particles.
In the Stokes' limit, the frequency of an isolated particle is given by $\omega_0=T/8\pi \mu R^3$. Using this, we calculate the rotational Reynolds number $\mathrm{Re}=\rho \omega_0 R^2/\mu=\rho T/8\pi\mu^2 R$ which is used in the text. The effective $\mathrm{Re}$ will be a little lower due to inertial effects~\cite{bickley1938lxv,liu2010wall,collins1955steady} and the hydrodynamic resistance from the wall.
We vary the particle area fraction $\phi_0 = N\frac{\pi R^2}{L_YL_Z}\times 100$\%, where $L_{Y|Z}$ are the simulation box lengths~\cite{SupMat}.
The hydrodynamic interactions create an attraction towards a no-slip surface along the spinning axis~\cite{liu2010wall}. To model an experimental set-up and to ensure a smooth monolayer we add an additional (weak) gravitational force towards the confining wall~\cite{SupMat}.

\begin{figure}
\centering
\includegraphics[width=1\columnwidth]{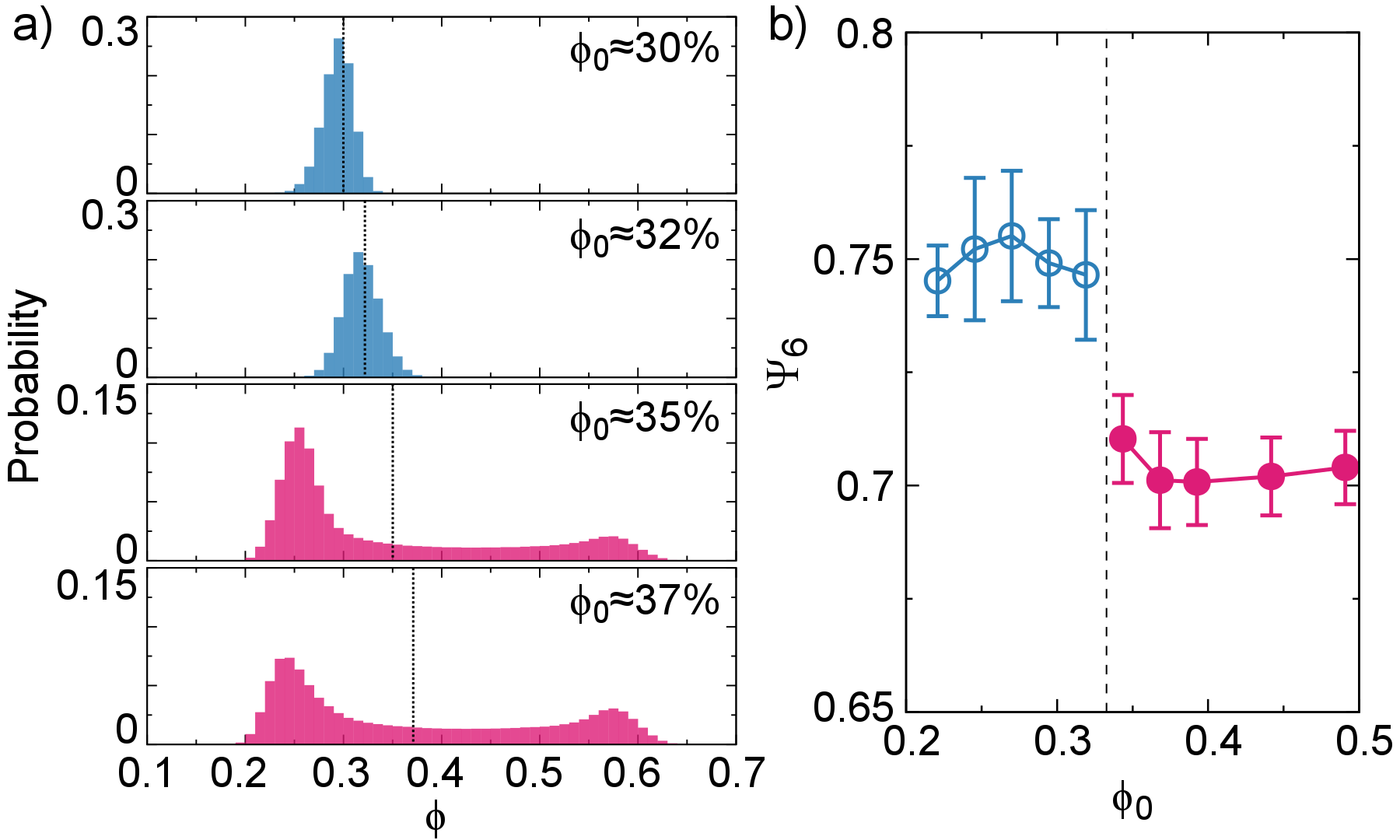}
\caption{(a) The probability distributions of the local area fractions $P(\phi)$ are shown for different global area fractions $\phi_0$ marked by the vertical dashed line for a $\mathrm{Re}\approx 10$ sample.  In the case of a single phase ($\phi<35\%$) the peak corresponds to $\phi_0$. The $P(\phi)$ becomes bimodal for $\phi_0 > 35 \%$. (b) The averaged local hexatic order as a function of overall area fraction.
\label{phase_phi}}
\end{figure}

\paragraph{Hexatic order and phase separation:}
{\color{black}A spherical particle spinning near a surface at $\mathrm{Re}>0$ creates an outward spiralling flow field (Fig.~\ref{config}(b))\cite{liu2010wall}. At higher particle area fractions this gives a rise to mixing via the rotating flow field, similarly to $\mathrm{Re} = 0$ case~\cite{lushi2015periodic,oppenheimer2019rotating} and it includes outward radial component, which leads to effective repulsions between the spinners.

When starting from random particle positions above the wall (Fig.~\ref{config}(a)), we observe a rapid formation of stable hexatic order of the spinners at semi-dilute area fractions (Fig.~\ref{config}(c)), which is in contrast with the $\mathrm{Re}=0$ case where the ordering requires a thermodynamic repulsion~\cite{oppenheimer2019rotating}.}  To characterise the ordered state, we calculate a local hexatic order parameter $\Psi_6$~\cite{SupMat}. A relative high value of $\Psi _{6} \approx 0.75$ (Fig.~\ref{phase_phi}(b) and Fig.~\ref{length}(a)) is observed when the overall area fraction $\phi_0 < 35\%$ {\color{black} and the local density distribution $P(\phi)$, calculated from sub-domains  $L_{Ys}=12R$ and $L_{Zs}=12R$, shows a single peak (Fig.~\ref{phase_phi}(a)).}

{\color{black} Increasing $\phi_0$, the $P(\phi)$ becomes bimodal (Fig.~\ref{phase_phi}(a)). The uniform structure deviates into a high density crystal surrounded by a lower density hexatically ordered state (Fig.~\ref{config}(d)). $P(\phi)$ shows two peaks at $\phi_1 \approx 25\%$ and $\phi_2 \approx 58\%$, corresponding to a low and a high particle density region, respectively (Fig.~\ref{phase_phi}(a)). The values of $\phi_1$ and $\phi_2$ are independent of the overall area fraction $\phi_0$, but the size of the dense region grows with the increasing $\phi_0$ while the size of the dilute region is reduced (as shown by the amplitudes of the two peaks in Fig.~\ref{phase_phi}(a)). Both the dense and dilute regions show a non-vanishing hexatic order,  with a slightly lower $\Psi _{6} \approx 0.7$ than in the single phase state (Fig.~\ref{phase_phi}(b) and Fig.~\ref{length}(a)).

\begin{figure}
\centering
\includegraphics[width=1\columnwidth]{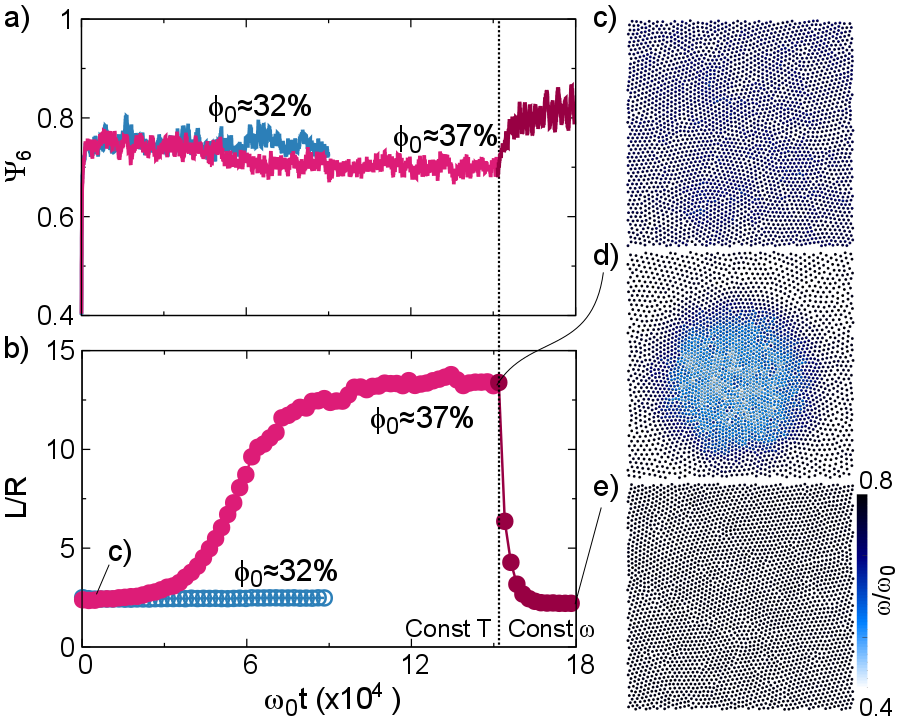}
\caption{The time evolution of {\color{black}(a) the local hexatic order parameter $\Psi_6$ and (b)} the domain length-scale $L$. When a constant torque was applied on each particle, the domain length shows a condensation of the particles for $\phi_0 \approx 37\%$ ($\mathrm{Re} \approx 10$). After the system reaches a steady two density state, the constant torque $T$ is replaced by a constant frequency $\omega$ corresponding to $\mathrm{Re}\approx 8$. {\color{black}(c-e)} The snapshots show particle positions and are color coded by the spinning frequencies $\omega$ of the particles.
\label{length}}
\end{figure}

To study the dynamics of the phase separation, we measure a time development of the domain length-scale $L(t)$~\cite{SupMat}. In the uniform state, the domain size is constant and {\color{black}the hexatic order grows rapidly (Fig.~\ref{length}(a) and (b))}. At higher area fractions, the local density becomes non-uniform and the domain length starts to grow (Fig.~\ref{length}(b),  see also Movie 1~\cite{SupMat}). After the onset, the growth of the $L(t)$ is rapid, eventually reaching a stable domain size. The particle spinning frequencies $\omega$  are strongly correlated with the local density. At high densities $\omega$ is decreased (Fig.~\ref{length}(d)), due to increased hydrodynamic resistance similar to what is observed with  passive colloidal particles~\cite{batchelor1972determination}. Replacing the constant torque $T$ by a constant spinning frequency, the two-density structure dissolves and a uniform hexatic state is reformed (Fig.~\ref{length}(b) and (e)).
These observations suggest that the density dependent spinning frequency can locally alter the hydrodynamic repulsion between the particles.

\textcolor{black}{Typical inertial (lift) forces on a spinning particle, such as Magnus effect, require that the object has a non-zero translational motion ~\cite{rubinow1961transverse}.
In our simulations, the particles have vanishing velocity $v << \omega R$ due to the constraint of the hexagonal crystal. We propose that the repulsion between the spinners mainly arises from the secondary flow created by the spinning spheres~\cite{climent2007dynamic,collins1955steady,bickley1938lxv}, and that the Magnus effect plays little or no role. For a single particle, the secondary flow has a radial component $v_r\sim \omega^2$~which advects the fluid away from the particle at the equatorial plane~\cite{climent2007dynamic,collins1955steady,bickley1938lxv}. Based on this, we expect that altering $\omega$ could lead to changes in the particle-particle interactions.}

\begin{figure}
\centering
\includegraphics[width=1\columnwidth]{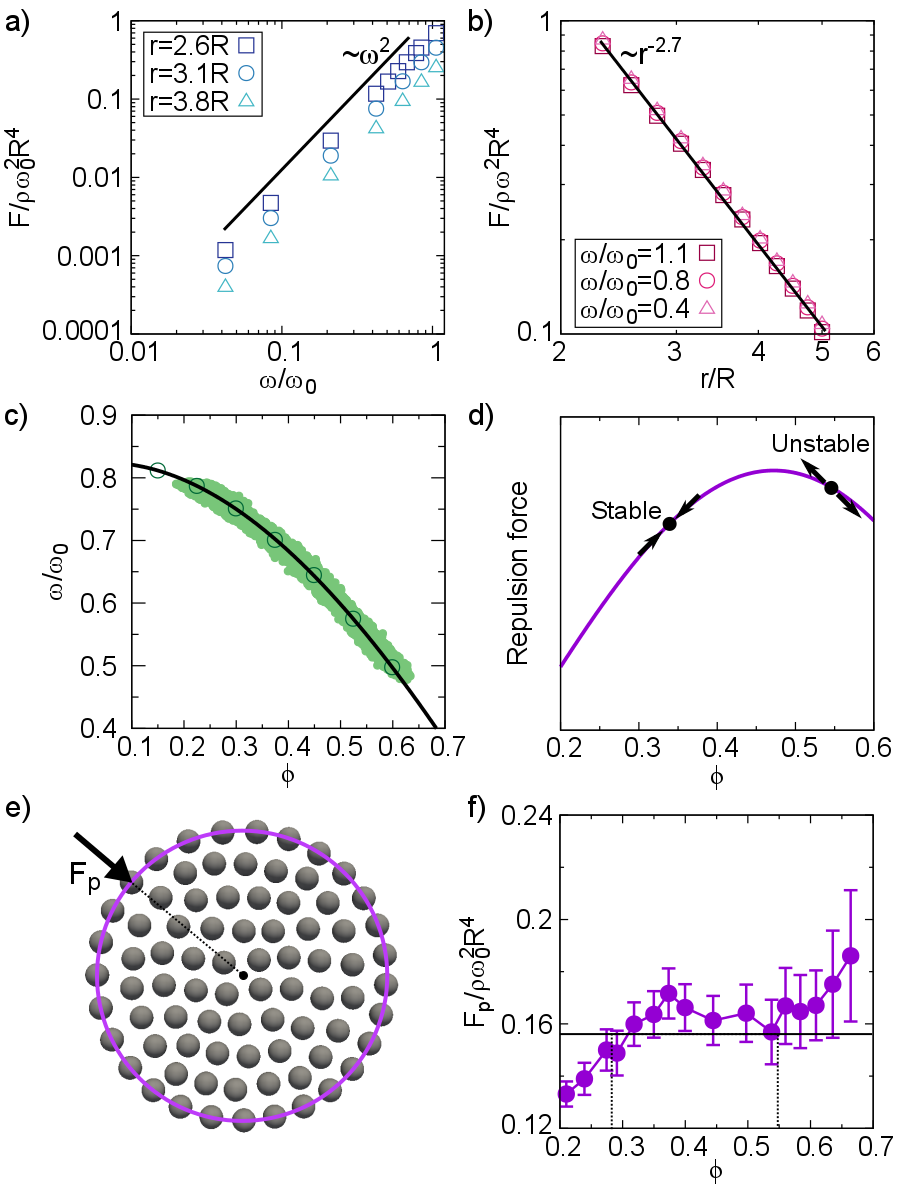}
\caption{(a, b) The hydrodynamic repulsion between a pair of spinners confined to move only along one direction ($Y$). (a) The repulsion force shows a $F \sim \omega ^2$ scaling and (b) a $F \sim r^{-2.7}$ decay. (c) The spinning frequency $\omega$ as a function of overall area fraction for a constant torque for a small sample  ({\color{black} open circles}) ($L_X=L_Y=L_Z=20R$ and $\mathrm{Re}\approx 10$). The black curve is a fit to polynomial function $\omega/\omega_0=0.308\phi^3-1.291\phi^2+0.122\phi+0.821$. {\color{black} The green shadow shows the correlation between the frequency and the local area fraction calculated from the two-phase state of Fig.~\ref{config}(d).} (d) A schematic showing the repulsion force as a function of overall area fraction when combing the measurements from (a-c). {\color{black}(e) A snapshot showing the measurement of the expansion force of a small spinner cluster ($N=80$). (f) The measured effective repulsion force of a spinner cluster as a function of the area fraction.}
\label{repulsion}}
\end{figure}

\paragraph{Non-monotonic repulsion and hydrodynamic instability:}

{\color{black}We estimate the repulsion force between a pair of spinners arising from the secondary flow by applying a spring force $F=-k_s(r-r_0)$ between the particles and varying $\omega$~\cite{SupMat}. The particles are restricted to a straight line along $Y$ to ensure that there is no translational motion.}
We observe a repulsive interaction $F\sim \omega^2$  between the  two spinners (Fig.~\ref{repulsion}(a)), which is in agreement with what is expected from the single particle flow field~\cite{climent2007dynamic,collins1955steady,bickley1938lxv}.}

{\color{black} For a single spinner the secondary flow shows a decay of the radial component as $v_r\sim \left(1-\frac{R}{r}\right)^2r^{-2}$~\cite{climent2007dynamic,collins1955steady,bickley1938lxv}. The interactions between the spinners is expected to be more complicated, due to the presence of the particle-wall and particle-particle near-field interactions.  For a particle pair, a decay $F\sim r^{-2.7}$ is observed and the normalised repulsion forces collapse on a single curve for all the spinning frequencies considered (Fig.~\ref{repulsion}(b)).}

Using a relation $r\sim \phi^{-0.5}$ between the area fraction $\phi$ and particle separation $r$ in a uniform two-dimensional structure and the data in Fig.~\ref{repulsion}(b), we can estimate a monotonic scaling of the repulsion force $F\sim\phi^{1.35}$ for a constant spinning frequency. This is not expected to lead to phase separation, in agreement with the constant $\omega$ case (Fig.~\ref{length}(e)).
When a constant torque is applied, the spinning frequencies are sensitive to the local surroundings and decrease when the local density is increased (Fig.~\ref{repulsion}(c), see also Fig.~\ref{length}(d)).

{\color{black} To qualitatively evaluate the existence of a critical area fraction $\phi^*$, we combine the pair data from Fig.~\ref{repulsion}a and b with the $\omega(\phi)$ data from Fig.~\ref{repulsion}c.  Now we can estimate $F\sim \omega^2(\phi) \phi^{1.35}$ which gives a non-monotonic $\phi$ dependence (Fig.~\ref{repulsion}(d)).}
At dilute regime the repulsion increases with increasing $\phi$ reaching a maximum at $\phi^* \sim 46\%$ and then starts to decrease (Fig.~\ref{repulsion}(d)). This can qualitatively explain our observation of the two density crystallisation. When the global area fraction $\phi_0 > \phi^*$, the uniform density is unstable, and any perturbation from the azimuthal mixing will lead to the separation to dense and dilute regions.
We note that our analysis in Fig.~\ref{repulsion}  is {\color{black}  based on pair interactions, and assumes a perfect symmetry.} It is over estimating the $\phi^*$ compared to the bulk simulations where $\phi^*\sim 35$\% is observed (Fig.~\ref{phase_phi}).  

{\color{black}
In the simulations, $\Psi_6$ is observed to grow rapidly to $\sim 0.75$, while the growth of the domain length-scale $L$ occurs at later stage (Fig.~\ref{length}(a) and (b)). When $\Psi_6 < 1$  the rotational (tangential) flow fields lead to translational motion of the spinners, giving a rise to non-uniformities in the particle density (see {\it e.g.} Movie 1 and 2 in~\cite{SupMat}). These suggest that local density fluctuations may reduce the $\phi^*$ predicted from the pair interactions in Fig.~\ref{repulsion}d. Close to the critical concentration, the density fluctuations would create short lived high density areas, eventually leading to the formation of a large high density cluster, stabilised by a boundary layer between the high and low density regions, where both the $\Psi_6$ and $\phi$ change continuously~\cite{SupMat} (Fig.~\ref{config}d). 


To better evaluate the effective repulsion between the spinners, we measure an expansion force of a small uniform spinner cluster~\cite{SupMat} (Fig.~\ref{repulsion}(e) and (f)).
The expansion force shows an increase until $\phi\sim 37$\%, and it is then observed to slightly decrease, followed by a steep increase at $\sim 55$\% due to particle collisions and repulsive lubrication forces (Fig.~\ref{repulsion}(f)).
This favours the formation of the  high density phase at $\sim 55$\%, in agreement with $\sim 58$\% observed in Fig.~\ref{phase_phi}. To balance the repulsion from the dense phase, we can estimate a low density phase at $\sim28$\% (Fig.~\ref{repulsion}(f)), which is close to $\sim25$\% observed in Fig.~\ref{phase_phi}.}

\paragraph{The effect of Reynolds number:}

\begin{figure}
\centering
\includegraphics[width=1\columnwidth]{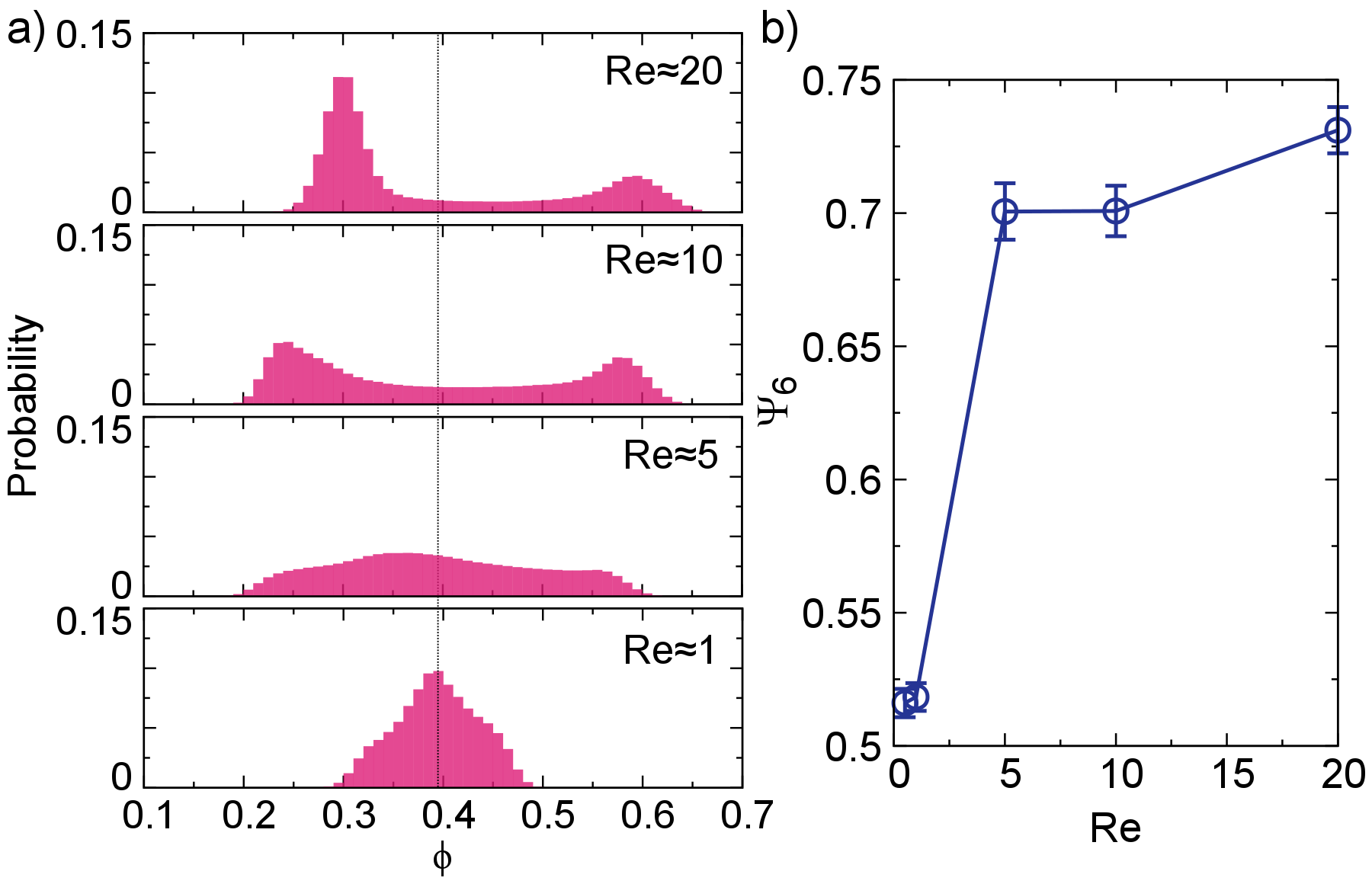}
\caption{(a) The probability distributions of the local area fractions $P(\phi)$ are shown for different Reynolds numbers for a global area fraction $\phi_0\approx 39$\% (marked by the vertical dashed line). (b) The averaged local hexagonal order parameter $\Psi_6$ as a function of the Reynolds number.
\label{phase_Re}}
\end{figure}

The inertial effects control the competition between the rotational mixing and the radial repulsion. For a single particle, the ratio between the azimuthal $v_\psi$ and radial $v_r$ flow fields gives $v_{\psi}/v_r\sim \mathrm{Re}^{-1}$~\cite{climent2007dynamic,collins1955steady,bickley1938lxv}. At small $\mathrm{Re}$ mixing dominates, and no spontaneous crystallisation is expected at the Stokes' limit in the absence of repulsive interactions~\cite{oppenheimer2019rotating}.

Starting from the steady state of a two-phase crystallisation ($\mathrm{Re}\approx 10$), we observe that the density peaks become less pronounced when $\mathrm{Re}$ decreases and disappears when $\mathrm{Re}\approx 1$ (Fig.~\ref{phase_Re}). Similarly, the hexatic order is lost when the relative hydrodynamic repulsion is reduced ($\mathrm{Re} < 5$ in Fig.~\ref{phase_Re}(b)){\color{black}, in agreement with the predictions at $\mathrm{Re}=0$~\cite{oppenheimer2019rotating}.}
When the $\mathrm{Re}$ is increased, the density peaks become more pronounced (Fig.~\ref{phase_Re}(a)) and the system shows increased hexatic order (Fig.~\ref{phase_Re}(b)).

\paragraph{Conclusions:}
We have simulated spherical particles spinning at an inertial regime. The results show that the particles form hexagonal structures when they settle on the solid surfaces with a semi-dilute particle concentrations.
Surprisingly, increasing the particle concentration, leads to a phase separated state, where the uniform hexagonal structure deviates into a dense domain surrounded by a less dense region while maintaining overall hexatic order. We demonstrate that this effect is due to the non-monotonic repulsion arising from the particle concentration dependent spinning frequency at  inertial regime.

We believe that our system can be useful for the design of new artificial spinner materials. {\color{black} A possible experimental realisation could be a spinner system consisting of millimeter sized particles  embedded with a weak magnet in a rotating magnetic field~\cite{godinez2012note}. By using weak magnets, and high magnetic field, the dipole-dipole interactions could be reduced enough to allow hydrodynamic interactions to dominate.}
Our observation of the phase separation provides a route for a plastic crystal state with a spatially variable density. Further it highlights the importance of inertial secondary flow in the spontaneous assembly of ordered structures in non-equilibrium spinner systems.


\begin{acknowledgments}
ZS and JSL acknowledge IdEx (Initiative d'Excellence) Bordeaux for funding, Curta cluster for computational time. JSL  acknowledges support by  the  French  National  Research  Agency  through  Contract No. ANR-19-CE06-0012-01.
\end{acknowledgments}

\bibliography{ref}


\pagebreak
\clearpage

\makeatletter 
\def\tagform@#1{\maketag@@@{(S\ignorespaces#1\unskip\@@italiccorr)}}
\makeatother

\makeatletter \renewcommand{\fnum@figure}
{\figurename~S\thefigure}
\makeatother

\setcounter{figure}{0}
\appendix
\section{Supplementary material}

\section{Simulation parameters}
For the Fig. 1, Fig. 2 and Fig. 3 in the main text, the simulations were carried out in a rectangular box $L_X=20R$ and $L_Y=L_Z=160R$ with no-slip walls at $X$ and periodic boundary conditions in $Y$ and $Z$. The parameters used in the simulation units (SU) are: the particle radius $R=4.1$, the viscosity $\mu=0.04$, the density of fluid $\rho=1$ and the torque $T=1.64$. These parameters give the Reynolds number $\mathrm{Re}=\frac{\rho \omega_0 R^2}{\mu}=\frac{\rho T}{8\pi\mu^2 R} \approx 10$. The additional gravitational force is $f_g =0.01$, which gives an isolated particle sedimentation velocity $u_s=\frac{f_g}{6\pi\mu R}\approx 0.033\omega_0 R$.

For the Fig.4 (a) and (b) in the main text, the simulations of pair interactions were carried out in a rectangular box $L_X=20R$, $L_Y=40R$ and $L_Z=20R$. The parameters are $R=4.1$, $\mu=0.04$ and $\rho=1$. The particles are set to have a constant spinning frequency. The particles are initialised on a straight line along $Y$ and a strong spring force $F_k=-k_s\Delta Z$ along $Z$ is applied. The spring constant $k_s$ is set to be a large enough value ($k_s=3000$ in simulation units) to ensure no rotation around the centre of mass of the pair ($\Delta Z \approx 0$). The repulsive force between the two particles, arising from the secondary flow, is measured by applying a spring force $F=-k_s(r-r_0)$ between the particles, and measuring $r$ for multiple $r_0$ using $k_s=3000$. The $r_0$ is the reference length and $r$ is the distance between the particles. 

For the Fig.4 (c) in the main text, the simulations were carried out in a rectangular box $L_X=L_Y=L_Z=20R$. The parameters are $R=4.1$, $\mu=0.04$, $\rho=1$ and $T=1.64$. The Reynolds number $\mathrm{Re}$ is about 10. The particle number $N$ is varied from 20 to 80 to get the desired area fractions.

For the Fig.5 in the main text, the simulations set-up is the same as the cases in the Fig.1. The viscosity $\mu$ and the torque $T$ are varied to obtain the desired Reynolds numbers.

\section{Measurement of the effective repulsion in a spinner cluster}
To estimate the effective repulsion in a spinner cluster, we measured an expansion force of a small particle cluster (Fig.4 (e) and (f) in the main text). The cluster consists of $N=80$ particles. The particles are initialised within a circular region with a radius $r_c$, which sets the particle fractions $\tfrac{NR^2}{r^2_c}$. The particles are driven by a constant torque, leading to a rotation at $\mathrm{Re}\approx 10$ as in Fig.1.

The cluster expands due to the inter-particle repulsion. We apply a confining spring force $F_p=-k_s(r-r_c)$ on the outermost layer of particles, when the distance $r$ of a particle centre is outside $r_c$. To reduce the particle tangential motion at the edge, a resistance $F_t=-k_t v_t$ is added opposed to the tangential velocity $v_t$ when the particle is outside of the circular region. At the steady state we obtain a circular spinner cluster with hexatic order ($\Psi_6 \sim 0.6$) as shown by Fig.4(e) in the main text. The time averaged $F_p$ is then measured and used to estimate the effective repulsion in the spinner cluster. The simulations are carried out with $k_s=30$ and $k_t=200$. The value of $r_c$ is varied to get the area fraction by $\phi=80 R^2/r_c^2$, and thus obtain the expansion force $F_p$ as a function of $\phi$ as shown in Fig.4(f) in the main text.

\section{Local hexatic order parameter}

\begin{figure}[h]
\centering
\includegraphics[width=0.5\textwidth]{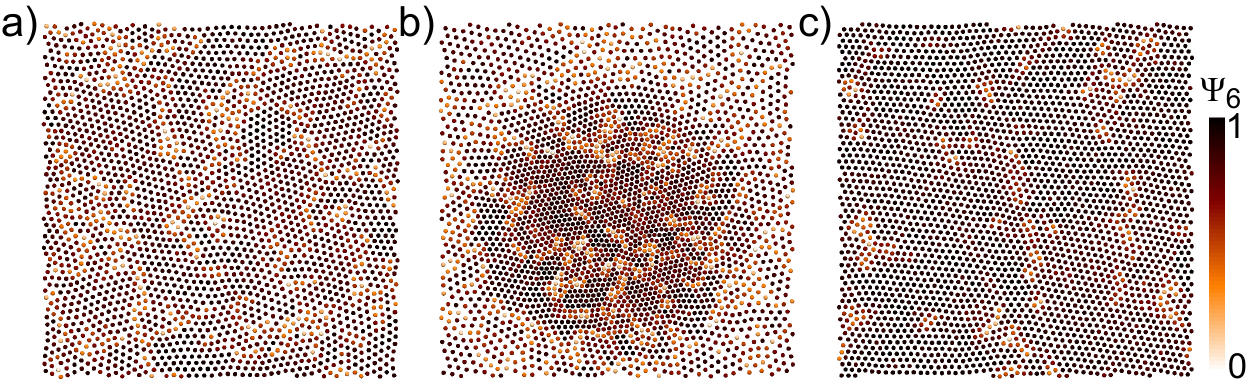}
\caption{
a) A snapshot for $\phi \approx 37 \%$ showing the configuration and local hexatic order for a uniform crystallisation at an early time. For $\phi \approx 37 \%$ the uniform state is unstable and deviates into a two-phase state, where the local hexatic order persists with a value of 0.7 but the local particle area fraction becomes a dense region surrounded by a dilute region as shown by b) a snapshot for $\phi \approx 37 \%$ in the steady state. When the constant torque is replaced by a constant frequency, the two-phase state dissolve and return to the uniform crystal with hexatic order parameter of 0.8 as shown by c) a snapshot for $\phi \approx 37 \%$ with constant frequency.}
\label{validation}
\end{figure}

We calculate a local hexagonal bond order parameter as  $\Psi _{6}=\left \langle  \left | \frac{1}{n_j}\sum_{k} \exp(i6\theta_{jk})  \right| \right \rangle$, where $n_j$ is the number of the nearest neighbours of particle $j$ and $\theta_{jk}$ is the angle between the vector connecting particles $j$ and $k$ and the horizontal axis ($Y$ in this work).

\section{Calculation of the domain length-scale $L(t)$}
To study the dynamics of the phase separation, we measure a time development of the domain length-scale  $L(t)=2\pi \int S(k,t)\mathrm{d}k/\int k S(k,t)\mathrm{d}k$,  defined as the inverse of the first moment of the spatially averaged structure factor $S(k,t)=\left \langle \phi(\textit{\textbf{k}},t)\phi(-\textit{\textbf{k}},t)\right \rangle$. The $\phi(\textit{\textbf{k}},t)$ is the spatial Fourier transform of the local area fraction and $k=\left | \textit{\textbf{k}} \right |$ is the modulus of the wave vector in Fourier space.

\end{document}